\documentclass[epj]{svjour}
\usepackage{latexsym}
\usepackage{graphicx}
\usepackage{amsmath,amssymb}
\newcommand{\nwc}{\newcommand}

\nwc{\ket}[1]{|#1\rangle}
\nwc{\bra}[1]{\langle#1|}
\nwc{\scal}[2]{\bra{#1}#2\rangle}
\nwc{\be}{\begin{equation}}
\nwc{\ee}{\end{equation}}
\nwc{\bea}{\begin{eqnarray}}
\nwc{\eea}{\end{eqnarray}}
\def\om{\omega}
\def\s{\sigma}
\def\O{\Omega}
\def\a{\alpha}

\def\la{\lambda}
\def\l{\left}
\def\r{\right}
\def\Neel{N{\'e}el\ }
\nwc{\bb}{\boldsymbol{\beta}}
\nwc{\ba}{\boldsymbol{\alpha}}
\nwc{\cA}{{\textsf{A}}}
\nwc{\cD}{{\textsf{D}}}
\nwc{\cQ}{{\textsf{Q}}}
\nwc{\cR}{{\textsf{R}}}
\nwc{\cS}{{\textsf{S}}}
\nwc{\cT}{{\textsf{T}}}
\nwc{\cV}{{\textsf{V}}}
\nwc{\Sv}{{\mathbf{S}}}
\nwc{\Hv}{{\mathbf{H}}}
\nwc{\ov}{\widehat{ {\mbox{\boldmath $\O$}} }}
\nwc{\fiv}{\widehat{ {\mbox{\boldmath $\phi$}}  }}
\nwc{\nv}{\widehat{ {\mbox{\boldmath $n$}}  }}
\nwc{\zv}{\widehat{ {\mbox{\boldmath $z$}}  }}
\nwc{\tS}{\widetilde{S}}
\nwc{\cH}{{\cal H}}
\nwc{\cG}{{\cal G}}

\begin{document}

\title{On $c=1$ critical phases in anisotropic spin-1 chains}


\author{C.~Degli Esposti Boschi\inst{1} \and E.~Ercolessi\inst{1,2} \and F.~Ortolani\inst{1,2} \and M.~Roncaglia
\inst{1,2}} 

\institute{INFM Research Unit of Bologna, \email{desposti@bo.infm.it}
\and
Physics Department, University of Bologna, and INFN, \email{ercolessi/ortolani/roncaglia@bo.infn.it}
\\6/2 v.le B.~Pichat, I-40127, Bologna, Italy.}

\abstract{Quantum spin-1 chains may develop massless phases in presence of Ising-like and single-ion anisotropies.  
We have studied $c=1$ critical phases by means of both analytical techniques, including a mapping of the lattice Hamiltonian onto an O(2) NL$\s$M, and a multi-target DMRG algorithm which allows for accurate calculation of excited states. We find excellent quantitative agreement with the theoretical predictions and conclude that a pure Gaussian model, without any orbifold construction, describes correctly the low-energy physics of these critical phases. 
This combined analysis indicates that the multicritical point at large single-ion anisotropy 
does not belong to the same universality class as the Takhtajan-Babujian Hamiltonian as claimed
in the past.
A link between string-order correlation functions and twisting vertex operators, along the $c=1$ line
that ends at this point, is also suggested.
}
\authorrunning{C. Degli Esposti Boschi et al.}
\titlerunning{On $c=1$ CFT for spin-1 chains}

\maketitle


\section{Introduction} \label{intro}

One-dimensional quantum spin systems have been extensively studied, since when \cite{Ha}, twenty years ago, Haldane argued that half-integer spin Heisenberg antiferromagnetic (AF) chains have no spin gap and are quantum critical, while integer spin chains are in the so-called Haldane gapped phase and exhibit correlation functions that decay exponentially with a finite correlation length. This scenario has then been confirmed both numerically \cite{WH} and experimentally \cite{Exp}. 

We will consider here instead the following spin-1 Hamiltonian
\begin{equation}
H=\sum_{j=1}^L \left\{S_j^x S_{j+1}^x + S_j^y S_{j+1}^y +
\lambda S_j^z S_{j+1}^z + D(S_j^z)^2 \right\} \; ,
\label{hamil}
\end{equation}
which includes both an Ising-like and a single-ion anisotropy term, with coefficients  $\la$ and $D$  respectively. In our calculations we impose periodic boundary conditions (PBC): $\Sv_{L+1}\equiv\Sv_{1}$. The inclusion of such terms is relevant for a better understanding of experimental compounds such as NENP and CsNiCl$_3$ (\cite{EMR} and references therein).

Even if no exact solution of (\ref{hamil}) is available (except for spin-1/2 \cite{KBI}), it is known from numerical studies \cite{BJ} that the inclusion of anysotropy terms can drive the system away from the Haldane phase toward other phases, some of which are critical. A first theoretical study which describes the critical properties of the model can  be found  in \cite{Sc} where, in general a spin-$S$ chain is mapped onto a system of $2S$ coupled spin-1/2 chains and analyzed with bosonization techniques. More recently, a thorough analytical study of this model has been presented in the seminal work \cite{KT}, where the emphasis is put on the physical properties of the massive phases.

The basic structure of the ground-state (GS) phase diagram of (\ref{hamil}) appears to be well understood \cite{CHS} and shows a rich variety of phases. For high values of $D$ the system is in the \textit{large-D phase} (D) consisting of a unique GS with total magnetization $S_{tot}^z=0$ separated by a gap from the first excited states which lie in the sectors $S_{tot}^z=\pm 1$.  For large positive values of $\la$ we have a twofold degenerate AF \textit{Ising-like phase} (I).  For not too large $D$ and $\la$, these two phases are separated by the \textit{Haldane phase} (H), which includes the isotropic O(3)-symmetric point. It is characterized by non vanishing string-order parameters
\begin{equation}
O_{S}^\alpha \equiv - \lim_{\vert j - k \vert \rightarrow \infty}  \left\langle S_j^\alpha \exp\l({ {\rm i} \pi \sum_{n=j+1}^{k-1} S_n^\alpha}\r) S_k^\alpha \right\rangle \; ,
\label{SOPjk}
\end{equation}
with $\a=\{x,y,z\}$, first introduced by den Nijs and Rommelse \cite{dNR}. The first excited states are magnons at the boundary of the Brillouin zone (BZ) carrying total spin 1. These three massive phases can be distinguished on the basis of a hidden $\mathbb{Z}_2\times \mathbb{Z}_2$ symmetry \cite{KT}, which is  fully (in the Haldane phase) or partially (in the AF Ising-like phase) broken and whose order parameters are given by Eq.  (\ref{SOPjk}). Also, the H-D and the H-I transition lines  meet at a tricritical point, for $D\simeq \la \simeq 3$. For larger values of the parameters, the Haldane phase disappears and the line $D\simeq \la$ represents a first order transition between the large-$D$ and the Ising-like phases. The remaining portion of the phase diagram, for $\la<0$, consists of a ferromagnetic GS for large  $|\la|$ and of 
{\it two critical gapless phases}, XY1 and XY2, different for having, for finite-size systems, first excitations carrying  $S_{tot}^z=\pm 1$ and $S_{tot}^z=\pm 2$ respectively.

In this paper, we will investigate the critical properties of the Hamiltonian (\ref{hamil}), and in particular we will concentrate on the Haldane/large-$D$ (H-D) transition line and on the XY2 massless phase, which turn out to be described by 
$c=1$ Conformal Field Theories (CFT). We will tackle the problem both via analytical tecniques, by finding the CFT Lagrangian that describes the model in the low-energy continuum limit,  and via a multi-target Density Matrix Renormalization Group (DMRG) algorithm, which allows for accurate calculation of excited states, to be compared with the operator content of the CFT. We will try also to clarify some controversial aspects discussed in the literature.

\section{The O(2) NL$\s$M on the H-D transition line } \label{sec:2}

The H-D transition line has been located numerically using the twisted boundary method in \cite{CHS}, where it has also been pointed out that this represents a second order phase transition described by a  $c=1$ CFT, in accordance with \cite{Sc}. 
Here we will describe  a mapping of the lattice model (\ref{hamil}) along the H-D transition line onto an O(2) nonlinear $\s$-model (NL$\s$M), that establishes a connection between the coupling constants $D, \la$ of the discrete model and those of the continuum Gaussian theory, namely the spin-wave velocity $v$ and the compactification radius. This will allow us to make quantitative predictions.

The partition function for Eq. (\ref{hamil}) in a path-integral representation which makes use of spin coherent states is given by \cite{Au}:
\begin{equation}
\mathcal{Z}=\int [D \ov]\exp\left\{ {\rm i} s \sum_{j}\om[\ov(j,\tau)]
-\int_0^{\beta}d\tau H(\tau)\right\}\; ,
\end{equation}
where the vector operator $\Sv_a(j)$ has been replaced by the classical variable $s\ov_a(j,\tau)$ and $\om[\ov_a(j,\tau)]$ is the Berry phase factor. In a semiclassical approach, we can expand $s\ov_a(j,\tau)$ about the classical solution which, for $D>\la-1$,  is a planar state where the the unit vectors $\ov_a(j,\tau)$ are \Neel ordered in the $xy$-plane:  $\ov(j,\tau)=(\cos(\theta_0 + j\pi), \sin(\theta_0 + j\pi),0)$. Hence we make the Haldane-like ans{\"a}tz:
\begin{equation}
\ov(j,\tau)=(-1)^{j}\nv(j,\tau) \sqrt{1-\frac{l^2(j,\tau)}{s^2}}
+ \zv\; \frac{l(j,\tau)}{s} \; ,
\label{ans}
\end{equation}
where $\nv(j,\tau)= {\rm e}^{{\rm i} \theta(j,\tau)}\in \mathrm{O(2)}_{xy}$, $\zv$ is the unitary vector
$(0,0,1)$, and the fluctuation field $l(j)$ is supposed to be small. Expanding  $H(\tau)$ up to quadratic terms in $l(j)$ and taking into account that the Berry phase is given by
\begin{equation}
{\rm i}s \sum_{j}\om[\ov(j,\tau)] = {\rm i} \sum_{j} \int_0^{\beta} d\tau\, l(j,\tau)\,
\partial_{\tau}\theta(j,\tau) \; ,
\label{top}
\end{equation}
we can now integrate out the fluctuating field. If we treat then $\theta(j,\tau)$ as a slow-varying variable, in the continuum limit we end up with an effective O(2) NL$\s$M in the field $\theta$ that, after the rescaling 
$\Theta = \theta/\sqrt{g}$, can be rewritten in the standard form:
\begin{equation}
\mathcal{L}_{O(2)}=\frac{1}{2}\left[\frac{1}{v}(\partial_{\tau}\Theta)^2+
v(\partial_{x}\Theta)^{2} \right]\; ,
\label{nlsm}
\end{equation}
where
\begin{equation}
g=\frac{1}{s}\sqrt{2\l(1+D+\la\r)}; \qquad v=s \sqrt{2\l(1+D+\la\r)}\; .
\label{gv}
\end{equation}
This is a free Gaussian model \cite{GNT}, with a bosonic field $\Theta$ compactified along a circle of radius 
$1/\sqrt{g}$, which describes a CFT with central charge $c=1$ and with primary fields (vertex operators) $V_{mn}$ of scaling dimensions given by
\begin{equation}
d_{mn}^{(th)}=\l( \frac{m^2 }{4 K}+n^2 K \r)\, , \qquad
m,n\in\mathbb{Z}\, ,
\label{d}
\end{equation}
where $K=\pi/g$ and
$n$ and $m$ are, respectively, 
the winding numbers of $\Theta$ and its dual field, $\Phi$. The latter turns out to be compactified
on a radius $\sqrt{g}/2\pi$. Moreover,
it is not difficult to see that $m=S_{tot}^z$, and hence it is a conserved quantity of the model. 
Indeed, from the NL$\s$M approach, it follows that $S^z(x) = l(x) = \partial_\tau \theta /vg$ 
so that
\begin{equation}
S^z_{tot} =\int dx \, \frac{\partial_{\tau}\Theta}{v\sqrt{g}}  =  \int dx \, \frac{\partial_{x} \Phi}{\sqrt{g}} 
= 2\pi m \frac{\sqrt{g}}{2\pi} \frac{1}{\sqrt{g}} = m \; .
\label{Sztot_vs_m}
\end{equation}
We recall that the values $K= \frac{1}{2}, 1,2$ correspond to the so-called \cite{Gi} self-dual (SD), 
free Dirac (FD) and Berezinski-Kosterlitz-Thosuless (BKT) points respectively.
The scaling  dimensions (\ref{d}) fix also the (non universal) critical exponents of the correlation functions. For instance the transverse spin-spin correlator decays according to:
\begin{equation}
\langle S^+(0) S^-(x) \rangle\approx \langle {\rm e}^{{\rm i}\theta(0)}  {\rm e}^{-{\rm i}\theta(x)} \rangle
\propto |x|^{-\eta}\; ,
\label{expo}
\end{equation}
where $\eta=2d_{10}=g/2\pi$.\\

Let us turn out to numerical results on the Hamiltonian (\ref{hamil}). We used a DMRG algorithm \cite{Wh} for finite systems, which had to be customized for the convergence of not only the GS but also of several excited state energies. 
More specifically, using the so-called thick-restart Lanczos algorithm \cite{WS} we could
target up to eight states in a given sector of $S^z_{tot}$. Then, the density matrix is built by averaging
(with equal weights) the density matrices associated with these states. Shortly, we will call this
a multi-target DMRG. If necessary, the correlation functions on the GS can be computed at the end
of the finite-size iterations (three in our cases).
Since the DMRG is known to be a good tool to investigate systems with a relatively short correlation length,
in order to study critical phases we had to handle the DMRG data with finite-size scaling (FSS) techniques. 
For the methodological aspects of our DMRG procedure, and fitting of data, 
we refer to a forthcoming paper \cite{CF}. 

First of all, it has been necessary to locate the H-D transition line with great precision. Initially, we have fixed some representative values of $\lambda$ and let $D$ vary across the phase boundary by small increments. Then, we have 
refined the location of the critical points $D_c(\lambda)$ according to FSS theory \cite{HB}, using $M=400$ DMRG states for chain lengths $L$ ranging from 10 to 50. The so obtained values are very close to the ones calculated in Ref. \cite{CHS} using the twisted boundary conditions method and exact diagonalization limited to $L=16$ sites.  
  
It is well known \cite{BCN} that in a CFT of a finite size system of length $L$, the GS energy density depends on the central charge and converges to its thermodynamic limit as 
\begin{equation}
\frac{E_{00}}{L}=e_\infty - \frac{\pi c v}{6 L^2} \; .
\label{E0_and_c}
\end{equation}
The excited state energies are instead related to the scaling dimensions (\ref{d}) by
\begin{equation}
E_{mn}-E_{00}= \frac{2\pi v}{L}\, \l( d_{mn} + r+ \bar{r} \r) \ , 
\label{levels}
\end{equation}
where $r$ and $\bar{r}$ are positive integers that label the secondary states of a Verma module \cite{DFMS}. 

\begin{figure}
\resizebox{0.5\textwidth}{!}{\includegraphics[keepaspectratio,clip]{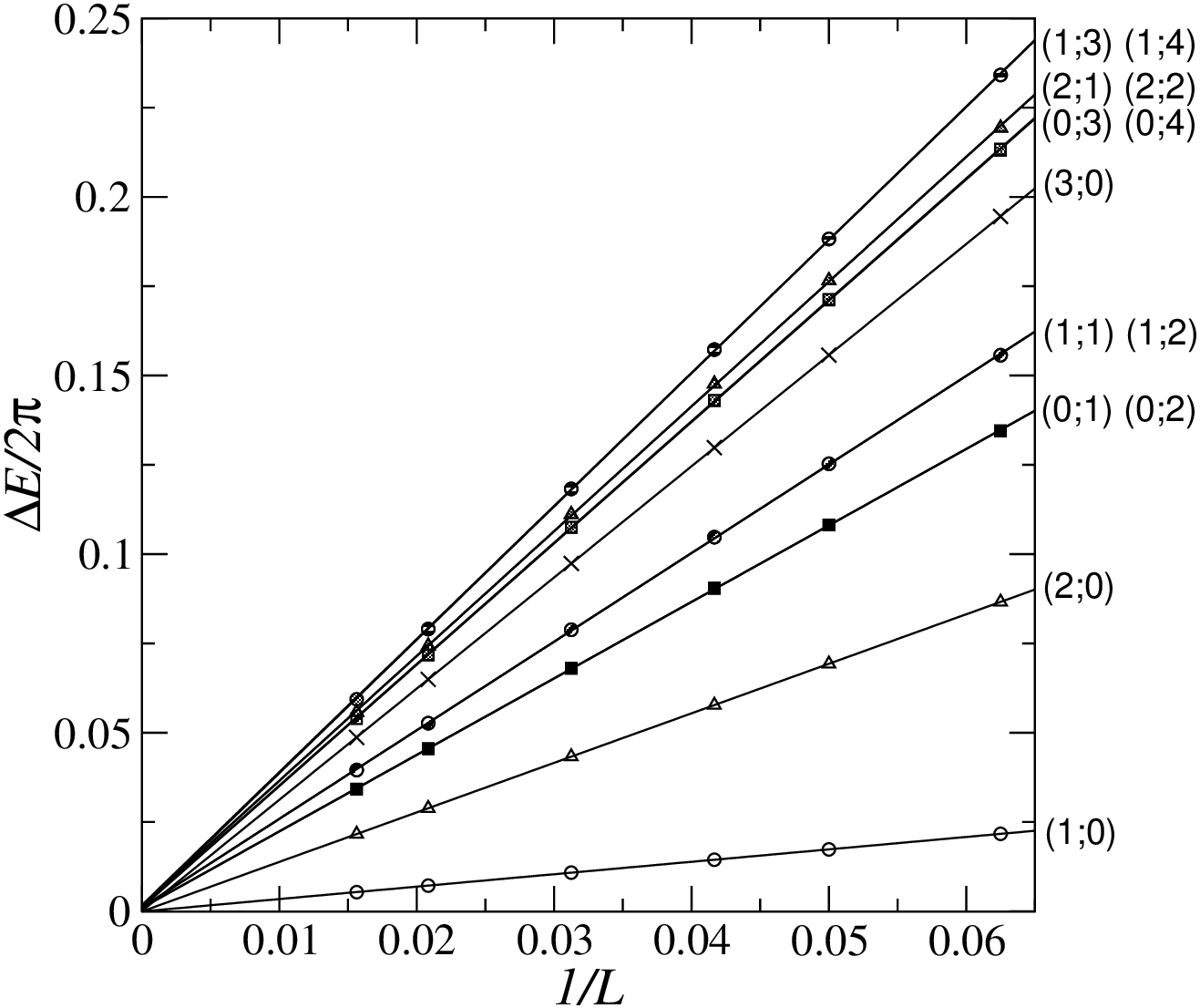}}
\caption{Energy differences, divided by $2 \pi$, plotted vs $1/L$ at the H-D transition point
$(\lambda=0.5,D=0.65)$. The legend on the right indicates the $S_{tot}^z$ quantum number and the
order of the targeted level within that sector, respectively. Points: multi-target DMRG data (with $M=405$ states).}
\label{ClHD065}
\end{figure}

In  a numerical approach, Eq.  (\ref{E0_and_c}) is the starting point to identify the correct CFT for a given critical point: accurate calculations of $E_{00}$ at various $L$  give in fact a best-fit of $e_\infty$ and of the  product $c v$. 
Then one has to select a number of excited states that become critical, i.e. degenerate with the GS, in the limit $L \rightarrow \infty$. In our problem the energies of the lowest states  have been calculated for different values of $S_{tot}^z$, which is the only quantum number that can be  fixed  within the DMRG algorithm.  If the hypothesis of an underlining $c=1$ Gaussian CFT is correct, all these energies should approach zero as straight lines as functions of $1/L$, according to Eq. (\ref{levels})  with $d_{mn}$ given in Eq.  (\ref{d}). 
For the critical point $(\lambda=0.5,D=0.65)$ this is shown in Fig.\ref{ClHD065}, where we have also indicated the 
(quasi-)degeneracy of each state, as observed numerically for chains of any length. The numerical data (points) of Fig.\ref{ClHD065} have been best-fitted with lines, whose slopes $d^{(num)}$ are reported in the second column of Table \ref{sdHD065}. 

In order to compare this spectrum with the theoretical predictions, it is necessary to find the values of $v^{(num)}$ and $K^{(num)}$, that identify the Gaussian model from which the observed levels originate.  Eq.  (\ref{d}) shows that, if $K>1$, the first excited state in the $m=0$ sector should be a doublet, corresponding to $m=n=0$ and $(r,\bar{r})=(1,0), (0,1)$. Numerically, we do indeed find a doublet as first excited states in  the $S_{tot}^z=0 $ sector, so that, using Eq. (\ref{levels}) together with the data that come from the fit on the GS energy of Eq. (\ref{E0_and_c}), we are able to obtain $v^{(num)}=2.197\pm0.004$ and $c^{(num)}=1.008\pm0.003$, which confirms the $c=1$ theory behavior.  
Also, again from  Eq.  (\ref{d}), it follows that the primary state ($m=1,n=0$) should be the lowest one  in the $S_{tot}^z=1$ sector,  with dimension $d_{10}=(4K^{(num)})^{-1}$. Our DMRG data yield then ${K}^{(num)}=1.580\pm0.004$
(\footnote{Throughout this paper the reported numerical errors originate from the best fits, with the
exception of the velocities for which the indicated errors represent the spread of the quadratic extrapolations
in $1/L$ of the secondaries of the GS.}).  
For comparison, if we plug the coordinates of the critical point $(\lambda=0.5,D=0.65)$  in the formulae (\ref{gv}) for $v$ and $g$ of the O(2) NL$\sigma$M we obtain $v^{(th)}=g^{(th)}=2.07$ and  $K^{(th)}=1.52$, which confirm the validity of our theoretical approach. 

We can now use the so calculated $v^{(num)}$ and $K^{(num)}$ to obtain the scaling dimensions $d^{(CFT)}$ of the low-lying levels as predicted by (\ref{levels}). These are listed in the first column of Table \ref{sdHD065} and are to be compared with the corresponding numerical observations given in the second column. One can see that the differences lie within 
a 2\% percent. As a final check of the validity of the NL$\s$M approach, we have computed directly the transverse spin-spin
correlation funcion, finding \cite{CF} that it decays algebraically with a critical exponent $\eta^{(num)}=0.312\pm 0.002$,
in very good agreement with the theoretical value $\eta=2 d_{10}=0.316$, obtained from  Eq.  (\ref{expo}).

\begin{table}
\centering{
\begin{tabular}{|l|c|c|r|}
\hline
$d^{(CFT)}$ [$\times$degeneracy] & $d^{(num)}$     & $(m,n)$   &  ($r,\bar{r}$)      \\
\hline
\hline
0 [$\times 1$]                        &                   & (0,0)       & (0,0)                   \\
\hline
0.1582 [$\times 2$] $^\dagger$       & 0.1582 $\pm$ 0.0004 & ($\pm 1$,0) & (0,0)                  \\
\hline
$0.633 \pm 0.002$ [$\times 2$]    & 0.631 $\pm$ 0.001  & ($\pm 2$,0) & (0,0)                      \\
\hline
1 [$\times 2$]                        & 0.975 $\pm$ 0.005 & (0,0)       & (1,0)              \\
                                      & 0.975 $\pm$ 0.005 & (0,0)       & (0,1)             \\
\hline
$1.1582 \pm 0.0004$ [$\times 4$] & 1.129 $\pm$ 0.006 & ($\pm 1$,0) & (1,0)                 \\
                                      & 1.129 $\pm$ 0.006 & ($\pm 1$,0) & (0,1)          \\
\hline
$1.424 \pm 0.004$ [$\times 2$]   & 1.416 $\pm$ 0.003 & ($\pm 3$,0) & (0,0)               \\
\hline
$1.580 \pm 0.004$ [$\times 2$]        & 1.546 $\pm$ 0.006 & (0,1)       & (0,0)              \\
                                      & 1.547 $\pm$ 0.006 & (0,$-1$)    & (0,0)            \\
\hline
$1.633 \pm 0.002$ [$\times 4$]  & 1.589 $\pm$ 0.007  & ($\pm 2$,0) & (1,0)                 \\
                                      & 1.589 $\pm$ 0.007   & ($\pm 2$,0) & (0,1)          \\
\hline
$1.738 \pm 0.004$ [$\times 4$]   & 1.693 $\pm$ 0.008 & ($\pm 1$,1) & (0,0)                    \\
                                      & 1.693 $\pm$ 0.008 & ($\pm 1,-1$) & (0,0)            \\ 
\hline                 
\end{tabular}
\caption{Spectrum of scaling dimensions at the point $(\lambda=0.5,D=0.65)$ on the H-D line, obtained from the scaling plots in Fig. \ref{ClHD065}. Here (H-D line) $m$ gives directly the eigenvalue of $S^z_{tot}$.
($^\dagger$ Setting $d^{(CFT)}=d^{(num)}$ fixes the value of $K^{(num)}=1.580$).}
\label{sdHD065}
}
\end{table}

We would like to stress here that our numerical analysis shows that the spectrum of the lattice Hamiltonian (\ref{hamil}) is completely exhausted by the levels of formula (\ref{levels}). This is at variance with some claims that have appeared in the literature in the past \cite{dNR} according to which the H-D transition line should be in the the same universality class as the  Ashkin-Teller (AT) model. Indeed it is known \cite{RY} that the critical properties of the latter are described by $c=1$ orbifold CFT models \cite{Gi} and hence should exhibit $K$-independent scaling dimensions as well.

\begin{figure}
\resizebox{0.5\textwidth}{!}{\includegraphics[keepaspectratio,clip]{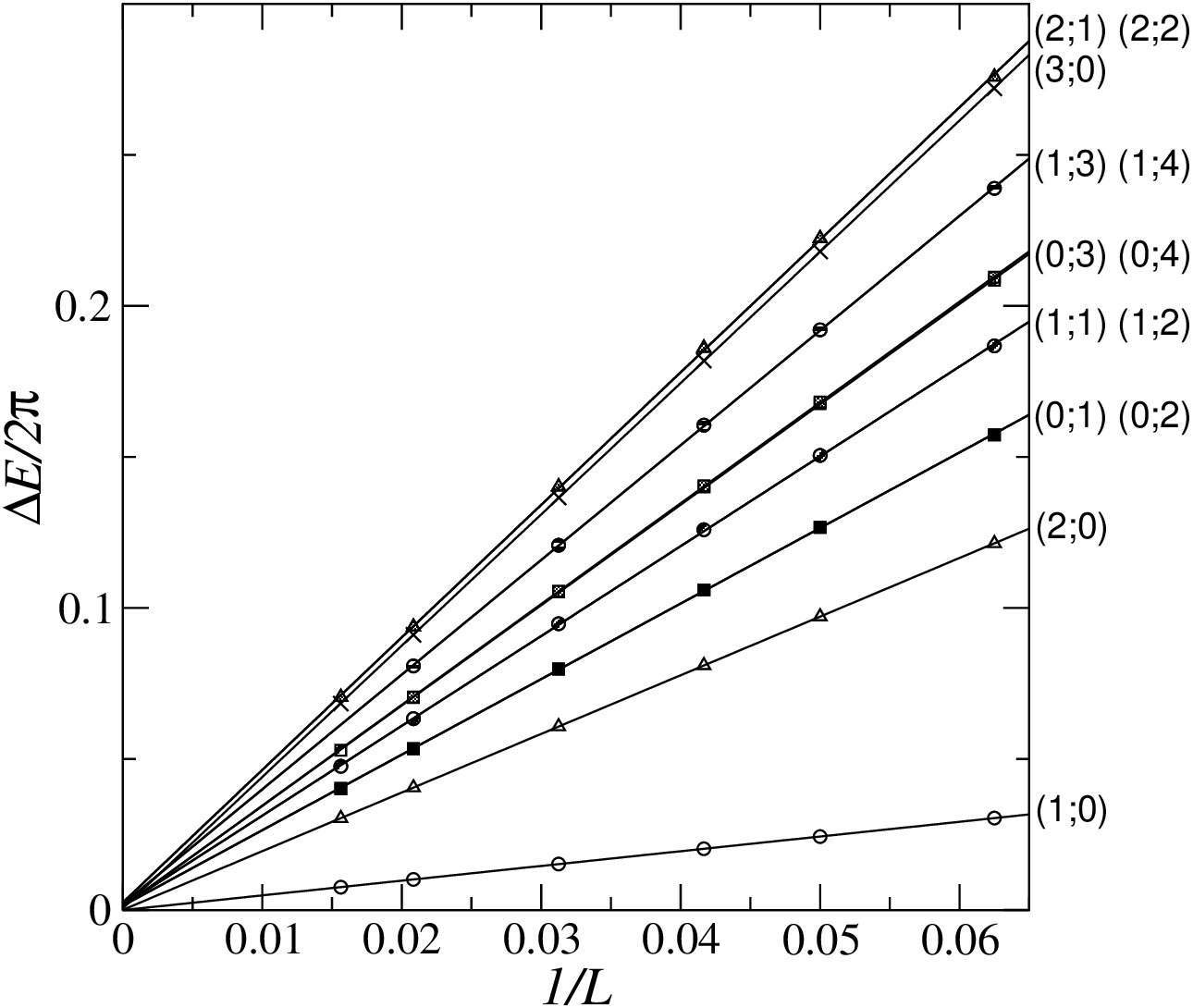}}
\caption{Energy differences, divided by $2 \pi$, plotted vs $1/L$ at the H-D transition point
$(\lambda=1,D=0.99)$. The legend on the right indicates the  $S_{tot}^z$ quantum number and the
order of the targeted level within that sector, respectively. Points: multi-target DMRG data
(with $M=405$ states). }

\label{ClHD099}

\end{figure}

This result can be further checked by  moving along the critical line, varying $K$. From a theoretical point of view this would correspond to acting by means of the marginal operator $(\partial_\mu \theta)^2$.  Consider for example the point $(\lambda=1,D=0.99)$, i.e. the $D$-induced transition point for the isotropic Heisenberg model. Numerically we have  estimated  $v^{(num)}=2.588\pm0.006$, $c^{(num)}=0.997\pm0.003$ and $K^{(num)}=1.328\pm0.004$. Again, we have  a good confirmation of the O(2) NL$\sigma$M predictions, corresponding to  $v^{(th)}=2.45$ and  $K^{(th)}=1.285$. For this case, scaling plots are displayed in Fig. \ref{ClHD099} and the scaling dimensions are listed in Table \ref{sdHD099}. Also, the calculated value of the critical exponent $\eta^{(num)}=0.374\pm 0.003$ matches the theoretical value $\eta=2 d_{10}=0.377$.     

\begin{table}[t]
\centering{
\begin{tabular}{|l|c|c|r|}
\hline
$d^{(CFT)}$ [$\times$degeneracy] & $d^{(num)}$       & $(m,n)$   &  ($r,\bar{r}$)       \\
\hline
\hline
0 [$\times 1$]                        &                     & (0,0)       & (0,0)                \\
\hline
0.1883 [$\times 2$] $^\dagger$       & 0.1883 $\pm$ 0.0005 & ($\pm 1$,0) & (0,0)                \\
\hline    
$0.753 \pm 0.002$ [$\times 2$]    & 0.750 $\pm$ 0.002    & ($\pm 2$,0) & (0,0)                \\
\hline
1 [$\times 2$]                        & 0.970 $\pm$ 0.006    & (0,0)       & (1,0)                \\
                                      & 0.966 $\pm$ 0.006    & (0,0)       & (0,1)                \\
\hline
$1.1883 \pm 0.0005$ [$\times 4$] & 1.148 $\pm$ 0.007   & ($\pm 1$,0) & (1,0)              \\
                                      & 1.148 $\pm$ 0.007   & ($\pm 1$,0) & (0,1)               \\
\hline
$1.328 \pm 0.004$ [$\times 2$]      & 1.284 $\pm$ 0.006   & (0,1)       & (0,0)                      \\
                                      & 1.289 $\pm$ 0.006   & (0,$-1$)    & (0,0)                      \\
\hline
$1.516 \pm 0.005$ [$\times 4$]& 1.468 $\pm$ 0.007   & ($\pm 1$,1) & (0,0)               \\
                                      & 1.468 $\pm$ 0.007   & ($\pm 1,-1$) & (0,0)              \\
\hline
$1.695 \pm 0.005$ [$\times 2$]   & 1.679 $\pm$ 0.005   & ($\pm 3$,0) & (0,0)                \\
\hline
$1.750 \pm 0.02$ [$\times 4]$   & 1.70 $\pm$ 0.01   & ($\pm 2$,0) & (1,0)                \\
                                      & 1.70 $\pm$ 0.01   & ($\pm 2$,0) & (0,1)               \\
\hline
\end{tabular}
\caption{Spectrum of scaling dimensions at the point $(\lambda=1,D=0.99)$, obtained from 
the scaling plots in Fig. \ref{ClHD099}. Recall that $m=S^z_{tot}$.
($^\dagger$ Setting $d^{(CFT)}=d^{(num)}$ fixes the value of $K^{(num)}=1.328$).}
\label{sdHD099}
}
\end{table}

\section{Towards the tricritical point} \label{sec:3}

As just recalled, some time ago it has been  conjectured \cite{dNR} that the H-D transition line is described by $c=1$ orbifold models. It has also been argued \cite{Sc} that the tricritical point at which the $c=1$ H-D transition line meets the $c=1/2$ H-I transition line is in the same universality class as the integrable Takhtajan-Babujian (TB) Hamiltonian
\cite{TB}, whose low-lying excitations are described by a SU(2)$_2$ Wess-Zumino-Novikov-Witten
(WZNW) model, with central charge $c=3/2$. Since this model is equivalent to three free relativistic Majorana fermions \cite{Ts}, one can conclude that the two critical lines must  meet at the exact point where the $c=1$  operator content is described by a bosonic theory with $K=1$ \cite{GNT}.
If so, moving towards the tricritical point on the H-D line one should find decreasing values of $K^{(num)}$ approaching 1 {\it from above}. Previous studies on the $\la$-$D$ phase diagram \cite{CHS} give $D\simeq \la \sim 3$ as a good estimate for the location of this point.

In the previous Section, we have seen that our NL$\s$M predictions seem to be quite reliable, at least for $D \lesssim 1$.
However, the assumptions under which  we have  derived the NL$\s$M should be more and more valid as $D$ (and $\la$) increases, since the true GS of the Hamiltonian (\ref{hamil}) is better and better described by the planar configurations of  Eq. (\ref{ans}). Tweaking the values of the microscopic parameters, we find that, in our approach, the condition $K=1$ is fulfilled for $\la\sim 2$. Thus our theoretical scenario predicts that the portion of the H-D line for $\la > 2$ must correspond to $K<1$. As we will see, our DMRG data confirm this hypothesis and at the same time give an estimate for $K$ at the tricritical point which is very close to the SD value $1/2$.

To check this numerically, we have considered the critical point $(\lambda=2.59,D=2.30)$. For such a point the analysis of the spectrum is more complicated than before because of two reasons. First, the H-D transition line is now very close to the H-I line and we start seeing a superposition of the two conformal spectra. However, the chosen point not being exactly on
the H-I transition line, from our numerical data we are still able to trace and separate the levels of the $c=1/2$ theory.
The price to be paid is that we had to target up to eight excited states within the $S^z_{tot}=0$ sector.
Second, since  we are moving towards the SD point, we have to take into consideration that the operators $V_{0\pm2}$, with scaling dimension $d_{0\pm2} = 4K$, are becoming less and less irrelevant (truly marginal for $K=1/2$). Thus, in a finite size system, we might expect \cite{CHS2} such operator to induce some renormalization effects on the energies of the model, which might be very large as we approach the SD point, where they are logarithmic. 
Another subtle point is the choice of the states from which the velocity is computed. When $K < 1$ we have to remember that the doublet with $m=0$ and $n=\pm1$ comes first
than the doublet of secondaries with $(r,\bar{r})=(1,0),(0,1)$. Even if we have no direct
control on how the aforementioned finite-size corrections act on this doublet, a quadratic extrapolation
in $1/L$ yields $v^{(num)}=3.70\pm0.04$, that seems to be quite reliable in that it gives $c^{(num)}=0.99\pm0.01$.
The theoretical value predicted by our NL$\sigma$M mapping is $v^{(th)}=3.43$.
On the other hand, in the estimate of $K^{(num)}$, the problem can be circumvented by using the dimensions
$d_{0\pm1}$ instead of $d_{10}$ as done before. In fact, the doublet with $m=0$ and $n=\pm 1$, which should be degenerate according to (\ref{d}), is found to be split. Following \cite{Ki}, we learn that the renormalization effects on the energies of these two states are equal but with opposite signs, so that the correct value of the energy can be obtained by considering their semisum. So doing, we get $K^{(num)}= 0.85 \pm 0.01$, confirming that we have gone closer to the SD point beyond the FD one. 

Finally, we have considered the point $(\lambda=3.20,D=2.90)$, which according to \cite{CHS} corresponds to the location of
the tricritical point. Similarly with what we have argued above, we find that the numerical spectrum can be interpreted as a superposition of a $c=1/2$ and a $c=1$ CFT's. The calculation of $c$ is complicated by the fact that the Ising
transition is even closer, if not coincident. Using the same method of the previous point,
our best values in this case are $c^{(num)}=1.133\pm0.006$ and $K^{(num)}=0.526\pm0.007$, having estimated
$v^{(num)}=4.445\pm0.005$ against $v^{(th)}=3.77$ from the expression of the NL$\sigma$M.
Interestingly enough, if we try to extract the Ising velocity from the secondaries
of the primary state with $d=1/8$ we find the pretty close value $v^{\rm Ising}=4.35\pm0.06$.

In Table \ref{vceiK_HD} we summarize the values of $v$ and $c$ for all the four H-D critical points discussed above and  we list also the GS energy per site $e_\infty$ and the final estimate of $K$. As anticipated, moving to
the right on the H-D line the value of $K$ keeps on decreasing towards the SD point where we
speculate that this line meets the H-I one and a first order transition line begins.

\begin{table}[h]
\centering{
\begin{tabular}{|c|c|c|c|c|c|}
\hline
$[\lambda,D_{\rm c}(\lambda)]$ & $v$ & $c$ & $e_\infty$ & $K$ & $\nu$ \\
\hline
\hline
$(0.5,0.65)$ & 2.197 & 1.008 &   $-0.908765(9)$  & 1.580 & 2.38 \\
$(1.0,0.99)$ & 2.588  & 0.997 & $-0.859152(2)$  & 1.328 & 1.49 \\
$(2.59,2.30)$ & 3.70  & 0.99 &   $-0.675099(5)$  & 0.85 & 0.870 \\
$(3.20,2.90)$ & 4.445 & 1.133 & $-0.59132(2)$ & 0.526 & 0.678 \\
\hline
\end{tabular}
\caption{Velocity, central charge and GS energy density (errors on the last figure in parenthesis) 
for the critical points discussed in the text. 
The numbers are the outcome of DMRG calculations with $L=16,20,24,32,48,64$ and $M=405$
for the first two lines and $L=16,20,24,28,32,36,40$ and $M=400$ for the other cases.
The last two columns contain, respectively, the estimate of the $K$ parameter of the effective $c=1$ Gaussian theory, according to the numerical procedure described in the text and the critical exponent $\nu=1/(2-K)$.}
\label{vceiK_HD}
}
\end{table}

\section{Away from the H-D critical line} \label{sec:4}

So far, we have considered what happens when moving along the $c=1$ critical line in the direction of increasing $\la$. When we instead move in the opposite direction, the velocity $v$ gets smaller and the compactification radius 
of $\Theta$ grows. In particular, when the H-D transition line meets the $\la=0$ axis, for $D \cong 0.4$
\cite{JP}, we reach a BKT 
transition, for which $K=2$. For $\la<0$ a two-dimensional critical region, corresponding to the XY1 phase, opens up. 

To understand what happens in the XY1 region as well as when we move away from the H-D line with $\la > 0$, we need to consider all relevant operators, allowed by the symmetries, that can be generated by renormalization of the lattice Hamiltonian. In a Gaussian theory with $1/2<K<2$, formula (\ref{d}) shows that the most relevant primary field not forbidden by the conservation of the total magnetization along the $z$-axis corresponds to $(m=0,n=\pm 1)$. It has scaling dimensions 
$d_{0\pm1}=K$ and it is given by the vertex operator $\cos{(\sqrt{4 \pi K}\Phi)}$. 

After a dual transformation on Eq.  (\ref{nlsm})
we find that our model has to be described by the Lagrangian of the  sine-Gordon model 
\begin{equation}
\mathcal{L}_{SG}=\frac{1}{2}\left[\frac{1}{v}(\partial_{\tau}\Phi)^2+
v(\partial_{x}\Phi)^{2} \right] + \frac{v\mu}{a^2} \cos{(\sqrt{4 \pi K}\Phi)} \; .
\label{sg}
\end{equation}
Here, we assume that the coefficient $\mu$ goes to zero along the H-D transition line. The precise determination of $\mu$ as a function of $\la$ and $D$ would require an exact renormalization procedure \cite{Sc,Ka}, which goes beyond the scope of this paper. The relevant cosine term is thus responsible of the opening of a gap as soon as we move away from the H-D transition line. In passing, we point out that, once the value of $d_{0\pm1}$ is known, the relation $\nu=1/(2-d_{0\pm1})$
yields a much better estimate, as compared to the $\beta$-function method \cite{CF}, of the exponent
that controls the opening of the energy gap proportional to $\vert D-D_c \vert^\nu$. The values computed with 
the former method are given in
the last column of Table \ref{vceiK_HD}.

The gap-generating term becomes marginal exactly at the BKT point $K=2$, ($\lambda=0,D\simeq 0.4)$
\cite{JP}, where the H-D line ``fans-out'' \cite{Ka} into the bidimensional critical region XY1.
We argue that in this critical phase the effective theory is the same as in Eq. (\ref{sg}), the
only difference being the irrelevance of the operator $\cos{(\sqrt{4 \pi K} \Phi)}$, with
$K$ ranging from 2, at the BKT boundary lines, towards $\infty$ at the ferromagnetic transition.
In order to support this picture, we note that the NL$\s$M approach predicts a critical stripe
enclosed between $D=-\lambda-1+\pi^2/8$ (where $K^{(th)}=2$) and $D=-\lambda-1$ (where $K^{(th)}=\infty$),
that overlaps with the real XY1 region in a wide portion of the diagram.
As a numerical test, we have considered the point $(\lambda=-0.5,D=0)$ where we expect $K^{(th)}=\pi$
and $v^{(th)}=1$. Using the same DMRG procedure as above ($L$ up to 40 with $M=400$), we find
that the low-lying spectrum is again described by a purely Gaussian CFT with $v^{(num)}= 1.11535\pm0.00005$,
$c^{(num)}=0.9997\pm0.0001$ and $K^{(num)}=3.086\pm0.002$.

To finish our discussion on the H-D transition, we observe that the operator content of the microscopic Hamiltonian (\ref{hamil}) does not contain the so-called twisting operator $V_{0\pm1/2} = \cos{(\sqrt{\pi K}\Phi)}$ with scaling dimension $d_{0\pm1/2} = K/4$. In general, this kind of operators is permitted in CFT \cite{Ki,Ka2}, in as much as they yield
half-integer conformal spin, and hence well-defined correlation functions \cite{GNT} decaying as power laws.
Specifically, for $V_{0\pm 1/2}$ the exponent has to be $\eta_{S,z}=2 d_{0\pm1/2}=K/2$.

This relevant operator is forbidden here only because of the PBC's we have chosen and not from symmetry considerations. 
Nonetheless, it is interesting to note that we {\it do} find certain correlation functions that, both
on analytical and numerical grounds, decay with the exponent written above.
Namely, we considered the longitudinal string correlator [$\alpha=z$ in Eq.  (\ref{SOPjk})] and extracted
its decaying exponent at the critical points of Table \ref{vceiK_HD} by means of a proper FSS analysis
\cite{CF,HB} on the DMRG data with $L=32,48,64,80,100$ and $M=300$.
We get $\eta_{S,z}^{(num)}=0.804 \pm 0.003$ at $(\lambda=0.5,D=0.65)$ and $\eta_{S,z}^{(num)}=0.741 \pm 0.002$
at $(\lambda=1,D=0.99)$, to be compared with $\eta_{S,z}^{(CFT)}=0.790$ and $\eta_{S,z}^{(CFT)}=0.664$,
respectively.
In addition, this preliminar identification is enforced by the continuum version of the string correlations
in the framework of the NL$\sigma$M. In fact, one can start from the local expression of $S^z$
on the H-D line
\begin{equation}
S^z(x) = \frac{\partial_x \Phi}{\sqrt{g}} + \kappa (-)^{x/a} \cos{(2 \sqrt{\pi K} \Phi)} + ...\; .
\label{Szx}
\end{equation}
with $\kappa$ a constant and $a$ the lattice spacing.
In the same spirit as bosonization, the first uniform term comes directly from
Eq.  (\ref{Sztot_vs_m}).
Since the Hamiltonian is quadratic in $S^z$, the second staggered term would generate
the first irrelevant cosine operator (allowed by symmetries) that one could put
in Eq.  (\ref{sg}) at criticality.
Now, independently of the scaling dimension of the exponential string in Eq.  (\ref{SOPjk}),
the derivatives that appear as a consequence of substituting Eq.  (\ref{Szx}) into the
outer spins $S^z_{j}$ and $S^z_{k}$ raise the overall scaling dimension by 1, and hence
cannot be responsible for the numbers above. So, the leading order is obtained
by retaining only the cosine term in the outer spin operators. 
Starting directly from the lattice formulation, it can be seen
that the alternating term $(-)^j$ cancels when combined with the exponential
of the string, in such a way that the sum in Eq.  (\ref{SOPjk}) can be effectively
taken only on the {\it uniform} part (defined as the average of the neighboring spins
in the doubled lattice cell). So in the continuum limit we are simply left
with $\int {\rm d}x \partial_x \Phi=\Phi(x)-\Phi(0)$. Summing up, we expect that the asymptotically dominant
contribution to the string correlator comes from
\begin{equation}
O_S^z(x) \approx \langle 0 \vert {\cal S^\dagger}(0) {\cal S}(x) \vert 0 \rangle \; ,
\;\; {\cal S}(x) \equiv {\rm e}^{ {\rm i}\frac{\pi}{\sqrt{g}} \Phi(x) } \cos{\left( 
\frac{2\pi}{\sqrt{g}} \Phi\right)} \; ,
\label{ce_SOPz}
\end{equation}
and particularly from the terms $\exp{(\pm{\rm i}\frac{\pi}{\sqrt{g}}\Phi)}$, 
that have the sought decay exponent $\eta_{S,z}=K/2$. Note that the prefactor $\pi$ in the string
of Eq.  (\ref{SOPjk}) is crucial for this.

Along the same lines, if we move off-criticality the first dominant term in Eq.  (\ref{Szx}) becomes
$(-)^{x/a} \cos{(\sqrt{\pi K }\Phi)}$ and, recalling that $K=\pi/g$, we see that Eq.  (\ref{ce_SOPz})
acquires a {\it constant} term due to the exponentials with opposite arguments.
We suspect that this is the mechanism according to which the string order parameters
become nonzero in the Haldane phase. Rigorously speaking, one should compute the off-critical correlators
using the sine-Gordon action and the outcome is expected to depend on the sign of $\mu$.
Here we can only notice that, semiclassically, when $\mu > 0$ the potential has a minimum in $\Phi_0=0$
while when $\mu < 0$ the minima lie at $\Phi_{\pm}=\pm \sqrt{\pi/4K}$.
The difference is that, in the first case $\cos{(\sqrt{\pi K }\Phi_0)} \neq 0$ while
in the second one $\cos{(\sqrt{\pi K }\Phi_{\pm})}=\cos(\pm \pi/2)=0$, which is the desired behavior.
However, a systematization of these ideas is still underway.

\section{The XY2 phase} \label{sec:5}

We consider now the XY2 phase, which coincides with a region of small negative $\la$ and $D\lesssim -2$, getting narrower as $D$ decreases. For large negative values of $D$ we can resort to a perturbative study, outlined in the Appendix, which  shows that the model is mapped onto an effective S=1/2 XXZ spin chain 
\begin{equation}\label{05xxz}
\cH^{\rm eff}= J \sum_{i=1}^{L}
\left(\tS_i^x \tS_{i+1}^x +\tS_i^y \tS_{i+1}^y +\Delta \tS_i^z \tS_{i+1}^z \right) \; ,
\end{equation}
where $J=1/|D|$, $\Delta=4\la |D|+1$ and $\mathbf{\tS}_i$ are spin-1/2 operators
(in particular $S^z_j= 2 \tS^z_j$).

From the exact solution of the  spin-$1/2$ XXZ model \cite{KBI}, we can therfore argue that our Hamiltonian, in the large $|D|$ limit, has a narrow critical region for $-(2|D|)^{-1}<\la\leq 0$, with elementary excitations carrying spin 1.  For positive $\la$ the system is in an AFM phase, while for $\la<-(2|D|)^{-1}$ it has a ferromagnetic GS. 
We can conclude that, in the continuum limit, the system can be mapped effectively onto a Gaussian model,  where the boson compactification radius and hence the critical exponents depend on $\Delta$. Also, at every point of the $\la$-$D$ parameter space there is a horizontal direction along which a marginal operator renormalizes the parameter $K$, 
and another direction  ($\la |D|=\mathrm{const.}$, for $D\ll -1$) along which the universality class does not change. 

\begin{figure}

\resizebox{0.5\textwidth}{!}{\includegraphics[keepaspectratio,clip]{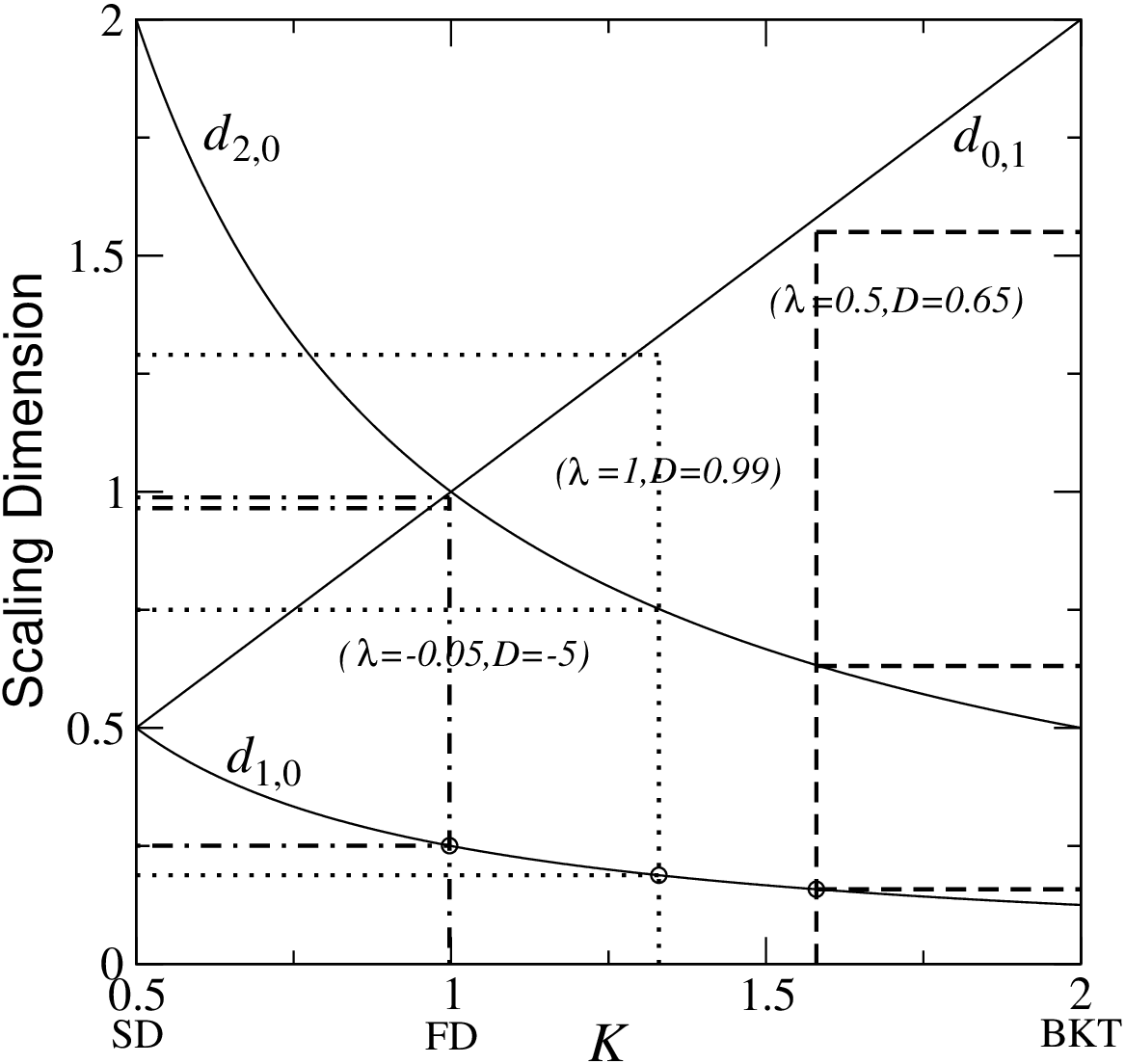}}

\caption{Scaling dimensions of the three most relevant primary operators, ($m=\pm 1, n=0$), ($m=0,n=\pm 1$) and ($m=\pm 2, n=0$) as functions of $K$. We have indicated the three special values  $K=1/2$: SD, $K=1$: FD; $K=2$: BKT). The vertical 
dot-dashed line indicate the value of $K=0.9976$ corresponding to the point $(\lambda=-0.05,D=-5)$ (in the XY2 phase) estimated through the first level $d_{10}$ (as indicated by the open circle).
The dotted and dashed lines have the same meaning for the points on the H-D line $(\lambda=1,D=0.99)$ and$(\lambda=0.5,D=0.65)$ respectively. In each case, the three horizontal lines mark the values of the direct numerical estimates of the
scaling dimensions.}

\label{SD_vs_K}

\end{figure}

Inspired by the above theoretical results, we have performed a multi-target DMRG calculation at the point $(\lambda=-0.05,D=-5)$, which in our mapping corresponds to a S=1/2 XXZ model with $\Delta=0$, described by a free Dirac fermion with velocity $v^{(th)}=1/\vert D \vert =0.2$. Consistently with theoretical predictions, we have found that only even spin sectors become gapless, while excitations with odd values of $S_{tot}^z$ remain massive in the thermodynamic limit. From the scaling of the GS and of the first doublet, we have obtained $e_\infty=-5.114607$, $v^{(num)}=0.20752\pm0.00006$ and 
$c^{(num)} = 1.01991\pm0.00003$. Keeping in mind that here $m$ has to be identified with twice the eigenvalue
of $S^z_{tot}$ we extract $K$ again from $d_{10}$, now associated with the first level in $S^z_{tot}=2$.
We find $K^{(num)}=0.9976 \pm 0.0004$, clearly compatible with the hypotesis of being at the FD point,
as seen in Fig. \ref{SD_vs_K} from the intersection of the dimensions $d_{0\pm1}$ and $d_{\pm20}$.

As for the spin-spin correlation functions, it turns out that the transverse ones, which do not have a direct identification in terms of spin-1/2 operators, have an exponential decay. On the other hand, at the free Dirac point the longitudinal spin-1 correlation function is simply $4\langle\tS_0^z \tS_j^z\rangle$ and sholud thearefore decay as $(1+(-1)^j)/j^{\eta_z}$, with $\eta_z=2$ \cite{GNT}. This alternating behavior is reproduced by DMRG calculations
($L=32,48,64,80,100$ with $M=300$) that yield the value $\eta_z=2.06\pm 0.03$.   

\section{Conclusions} \label{sec:6}

In this paper we have studied the spin-1 AFM Heisenberg chain with the inclusion of both Ising-like and single-ion anisotropy terms with the aim of investigating the critical properties of this model. In particular we have examined the massless phases that correspond to a $c=1$ CFT.

Our analysis starts from an analytical approach, aimed at the identification of the effective continuum field 
theory that describes the low-energy sector of the lattice model. 
We have found that all $c=1$ phases are described by a free Gaussian model (with no orbifold construction), with continuously varying critical exponents. These results have been then checked numerically, with the use of a multi-target DMRG algorithm, which allows for accurate calculation of many excited states together with string and ordinary correlation
functions. The agreement between the two methods is remarkable. 

In particular, along the H-D transition line both the NL$\sigma$M predictions and the DMRG data indicate
that the Gaussian parameter $K$ changes continuously from the value $K=2$ (corresponding to the BKT point) for $\la=0$ to $K=1/2$ (corresponding to the SD point) for the tricritical point where the H-D and the H-I transition lines meet. The latter result is at variance with some claims \cite{Sc,dNR} according to which this point should be described by an 
SU(2)$_2$ WZWN model. Moving away from the H-D line, a gap opens up due to the relevance of
the cosine operator in Eq. (\ref{sg}). An analytical argument, supported by numerical estimates of the decay exponents,
that relates the so-called twisting operator to the longitudinal string order correlation function is also
skecthed. 
As far as the XY1 phase is concerned, there are analytical and numerical evidences that the effective CFT 
is again a Gaussian model with $K > 2$. In this case the sine-Gordon operator is irrelevant.

Finally, it is shown that, in the XY2 phase, the low-energy physics of the spin-1 model is equivalent to a XXZ spin-1/2 chain with an anisotropy parameter $-1< \Delta \le 1$.


\begin{acknowledgement}
We would like to thank L. Campos-Venuti, G. Morandi, S. Pasini and F. Ravanini for useful discussions. This work has been in part supported by the TMR network EUCLID (contract number: HPRN-CT-2002-00325).
\end{acknowledgement}



\section*{Appendix: Large Negative $D$}\label{sec:7}

For convenience, we consider the Hamiltonian (\ref{hamil}) subtracting out a constant term $DL$.
in order to study the $D\to - \infty$ limit, keeping $\la$ finite.
At the zero order in  perturbation theory we simply have the Hamiltonian 
\begin{equation}
\cH^0 = -|D| \sum_{j=1}^{L} \left(S_j^z\right)^2 + |D|L \; ,
\end{equation}
whose GS has degeneracy $2^L$, corresponding to all the spin configurations with $S_j^z = \pm1$, i.e. not containing zero's. 
This degeneration is lifted by the perturbation part
\begin{equation}
\cH^1=\sum_{i=1}^{L}\left[\frac{1}{2}\left(S_i^+ S_{i+1}^- +S_i^- S_{i+1}^+ \right)
+ \la  S_i^z S_{i+1}^z \right] \; . 
\end{equation}

We denote with  $P_0$ and $P_1$  the projectors onto the subspaces
$Z^0=\left\{|s_1,s_2,\cdots,s_L\rangle : \quad \forall i,
|s_i\rangle \not=|0\rangle \right\} $ and $Z^1=\left\{|s_1,s_2,\cdots,s_L\rangle : \quad \exists i, |s_i\rangle= |0\rangle \right\} $ respectively, so that we can write:
\begin{equation}
\cH=\l(
\begin{array}{cc}
P_0\cH^1 P_0\ & \ P_0\cH^1 P_1 \\
P_1\cH^1 P_0\ & \ P_1(\cH^0+\cH^1)P_1 
\end{array} \r) \; .
\end{equation}
We can now look for  an effective Hamiltonian $\cH^{\rm eff}$ describing  the low energy sector by projecting the resolvent operator $\cG(E)$ onto the subspace $Z^0$ \cite{Au}:
\begin{equation}                                                     \label{res}
P_0\cG(E) P_0=P_0[E-\cH]^{-1} P_0 \equiv [E-\cH^{\rm eff}(E)]^{-1} \; .  
\end{equation}
We have
\begin{equation}\label{eff1}
\cH^{\rm eff}=P_0\cH^1 P_0 + P_0\cH^1 P_1
\{P_1[E-(\cH^0+\cH^1)]P_1\}^{-1} P_1 \cH^1 P_0
\end{equation}
where we can consider the  approximation in which $E=0$ and expand to the second order in $\cH^1$. Since the off-diagonal part of $\cH^1$ connects $Z^0$ to $Z^1$ creating a couple of zeroes, at the leading order we find
\begin{eqnarray}\label{eff2}
\cH^{\rm eff}&=&P_0\cH^1 P_0 - \frac{1}{2|D|} P_0\cH^1 P_1 \cH^1 P_0  \label{eff3}\\
&=& \sum_{j=1}^{L} \left[ \la S_j^z S_{j+1}^z  - \frac{1}{8|D|}
\left(S_j^+ S_{j+1}^- +S_j^- S_{j+1}^+ \right)^2\right] \; . \nonumber
\end{eqnarray}
This effective Hamiltonian is acting within $Z^0$ where we have only two local states per site. So, within this subspace, we get an identification of the local spin-1 operators $S_j^\alpha$ with spin-1/2 operators $\tS_j^\alpha$ according to the following table.
\begin{center}
\begin{tabular}{|c|c|}
\hline
 Spin 1        & Spin 1/2 \\
\hline
\, $S^z_j$ \, & $2 \tS^z_j$   \\
\, $S^+_j S^+_j$ \, & \, $2 \tS^+_j$ \,       \\
\, $S^-_j S^-_j$ \, & \, $2 \tS^-_j$ \,     \\
\, $S^+_j S^-_j$ \, & \, $2 \left(\frac{1}{2}+\tS^z_j\right)$ \,       \\
\, $S^-_j S^+_j$ \, & \, $2 \left(\frac{1}{2}-\tS^z_j\right)$ \,       \\
\hline
\end{tabular}
\end{center}

In terms of the new operators, we obtain an effective spin-1/2 XXZ model
\begin{eqnarray}\label{eff4}
\cH^{\rm eff}&=& \frac{1}{|D|} \sum_{j=1}^{L}
\left(\tS_j^x \tS_{j+1}^x +\tS_j^y \tS_{j+1}^y + 
(4\la |D|+1)\tS_j^z \tS_{j+1}^z \right)\nonumber \\ 
&&\qquad \qquad -\frac{N}{4|D|} \; , 
\end{eqnarray}
where the sign in front of the $x$-$y$ terms has been changed by means of the unitary
transformation $\tS_j^x \to (-)^j \tS_j^x$, $\tS_j^y \to (-)^j \tS_j^y$.


\end{document}